\documentclass[sort&compress,final,3p,times,twocolumn,fixfloat]{elsarticle}
\usepackage{comment}
\usepackage{graphicx,color}
\usepackage[T1]{fontenc}
\usepackage{dcolumn}
\usepackage{bm}
\usepackage{caption}
\usepackage{amsfonts}
\usepackage{multirow}
\usepackage{mathrsfs}
\usepackage{amsmath}
\usepackage{amssymb}
\usepackage{bbm}
\usepackage{color}
\usepackage{array}
\usepackage{diagbox}
\usepackage{float}
\usepackage{txfonts}
\usepackage{bm}
\usepackage{booktabs}
\usepackage{comment}
\usepackage{lineno}
\usepackage{xspace} 
\usepackage{scalerel}
\usepackage{tikz}
\usepackage{cuted}

\def\be{\begin{equation}}
\def\ee{\end{equation}}
\def\bea{\begin{eqnarray}}
\def\eea{\end{eqnarray}}

\allowdisplaybreaks[2]

\begin{document}

\title{Decay constant of $B_c^*$ accurate up to $\mathcal{O}(\alpha_s^3)$}

\date{today}

\address[a]{School of Physical Science and Technology, Southwest University, Chongqing 400700, China\vspace{0.2cm}}
\address[b]{College of Big Data Statistics, Guizhou University of Finance and Economics, Guiyang, 550025, China}

\author[a]{Wen-Long Sang}
\ead{wlsang@swu.edu.cn}
\author[b]{Hong-Fei Zhang}
\ead{shckm2686@163.com}
\author[a]{Ming-Zhen Zhou}
\ead{zhoumz@swu.edu.cn}

\begin{abstract}
In this paper, we evaluate, up to QCD next-to-next-to-next-to-leading order, the $B_c^*$ decay constant,
which, within the nonrelativistic QCD (NRQCD) framework,
is factorized as the product of the short-distance coefficient (SDC) and the long-distance matrix element.
For the first time, the renormalization constant and the anomalous dimension for the NRQCD vector current composed of $c\bar{b}$,
which are functions of the charm quark mass $m_c$, the bottom quark mass $m_b$ and the factorization scale
$\mu_\Lambda$,
are obtained analytically at $\mathcal{O}(\alpha_s^2)$ and $\mathcal{O}(\alpha_s^3)$.
The SDC is calculated up to
$\mathcal{O}(\alpha_s^3)$ in perturbative expansion.
Explicitly, the $\mathcal{O}(\alpha_s^2)$ correction to the SDC is analytically calculated in terms of logarithmic and polylogarithmic functions of $r\equiv m_b/m_c$,
and the $\mathcal{O}(\alpha_s^3)$ correction to the SDC is numerically calculated at a series of values of $r$, ranging from $2.1$ to $4.0$.
Surprisingly, we find that the nontrivial part of the SDC at $\mathcal{O}(\alpha_s^3)$ can be well estimated by a linear function of $r$.
In addition, We find that the $\mathcal{O}(\alpha_s^2)$ and
$\mathcal{O}(\alpha_s^3)$ corrections to the decay constant and decay width are considerable and very significant, which indicates a very poor convergence for the perturbative expansion.
\end{abstract}

\maketitle

\section{Introduction}

The $B_c$ meson was first discovered by \texttt{CDF}~\cite{CDF:1998ihx,CDF:1998axz} at the Fermilab Tevatron through the semi-leptonic decay $B_c\to J/\psi+l+\nu_l$ at the end of last century.
Soon afterwards, more data about this particle were measured in experiments~\cite{CDF:2005yjh, CDF:2007umr, D0:2008bqs, D0:2008thm, Corcoran:2005ti}.
The first excited state $B_c(2S)$ was discovered by \texttt{ATLAS} through its hadronic transition to the ground state~\cite{ATLAS:2014lga},
and was confirmed a few years later by \texttt{CMS}~\cite{CMS:2019uhm}.
In 2019, the observation of the vector $B_c(2S)^*$ through its decay into $J/\psi\pi\pi$ was reported by \texttt{CMS} and \texttt{LHCb}~\cite{CMS:2019uhm,LHCb:2019bem}.
However, the ground state $B_c^*$ has not been observed yet.
It is expected that there will be a lot of $B_c^*$ events corresponding to a dataset of 300 $\rm fb^{-1}$ at LHCb,
therefore, it is required that the processes, $B_c^*\to\tau\nu_\tau$, $B_c^*\to\mu\nu_\mu$,
and $B_c^*\to e\nu_e$ be scrutinized from both theoretical and experimental sides.

In the past several decades, a lot of efforts~\cite{Chang:1992pt, Braaten:1995ej, Chang:2000ac, Chang:2001pm,Chang:2005wd,Wang:2015bka,Zhu:2017lqu,Chen:2020ecu,Sun:2022hyk,Wang:2012kw}
have been made to unravel the nature of this particular quarkonium.
Unlike the vector charmonium and bottomonium,  which can decay into lepton pair through the electromagnetic current,
the decay of the $B_c^*$ meson can proceed only through the charged weak current.
The $B_c^*$ decay constant is a universal parameter which can be conveniently employed in the calculation of the corresponding weak decay processes.
Thanks to the nonrelativistic feature of the charm and bottom quark in the $B_c^*$ meson,
this decay constant can be perturbatively evaluated in powers of $v_Q$ and $\alpha_s$ in the NRQCD~\cite{Bodwin:1994jh} scheme.
The $\mathcal{O}(\alpha_s)$ radiative correction and $\mathcal{O}(v^2)$ relativistic correction have been computed in Ref.~\cite{Hwang:1999fc}.
The $\mathcal{O}(\alpha_s v_Q^2)$ correction, with partial high-order relativistic corrections resummed, has been investigated in Ref.~\cite{Lee:2010ts}.

During the past two decades, we have continuously witnessed the progress in computing the three-loop corrections to quarkonium decays.
The results of the three-loop QCD corrections to the decay constant of $J/\psi$/$\Upsilon$ have been given in Refs.~\cite{Marquard:2006qi,Marquard:2009bj,Marquard:2014pea,Beneke:2014qea,Egner:2022jot}.
The missing singlet contribution and the double heavy quark mass effect were obtained very recently in Ref.~\cite{Feng:2022vvk}.
The challenging three-loop QCD correction to the decay constant of $B_c$ was computed in Ref.~\cite{Feng:2022ruy},
where both the charm-quark and bottom-quark loops are counted.
In this work, we consider the three-loop QCD correction to the decay constant of the vector $B_c^*$ meson.

The rest of this paper is organized as follows. In Sec.~\ref{sec:general},
we employ the NRQCD formalism to factorize the decay constant, present the framework for computing the SDC,
and analyze the general structure of the SDC.
In Sec.~\ref{sec:tech}, we describe the techniques used in our computation.
In Sec.~\ref{sec:two-loop}, we present the analytical expressions of the SDC at one-loop and two-loop orders.
We devote Sec.~\ref{sec:three-loop} to the three-loop correction to the SDC.
The renormalization constant and anomalous dimension associated with the NRQCD vector current is also obtained.
Section~\ref{sec:phy} is devoted to the phenomenological analysis and discussion.
A summary is given in Sec.~\ref{sec:sum}.
In Appendix~\ref{sec:app}, we present the numerical results of the SDC at various values of $r$.

\section{General description~\label{sec:general}}

The decay constant $f_{V}$ of $B_c^*$ can be related to the vacuum-to-$B_c^*$ matrix element mediated by the QCD vector current:
\bea\label{eqn:dc full}
\langle 0| \bar{b} \gamma^{\mu} c |V \rangle= -  f_{V}m_V \epsilon_V^{\mu},
\eea
where $V$ represents $B_c^*$, which is
relativistically normalized, $m_V$ and $\epsilon_V$ denote the mass and polarization of $V$, respectively.

Since both the charm and bottom quarks move nonrelativistically inside $V$, we can employ the NRQCD framework to compute
$f_V$.  According to the NRQCD factorization formalism, we have
\bea\label{eqn:fac}
&&f_{V}= \sqrt{\frac{2}{m_{V}}}\, \mathcal{C}(m_c,m_b,\mu_{\Lambda}) \nonumber \\
&&~~~~\times\langle 0 |\chi^{\dagger}_b\bm{\sigma}\cdot \bm{\epsilon}\psi_c(\mu_{\Lambda})|V \rangle+\mathcal{O}(v^{2}),
\eea
where $\chi^{\dagger}_b$ and $\psi_c$ denote the Pauli spinor fields annihilating the $\bar{b}$ and $c$ quarks, respectively,
$\mathcal{C}$ denotes the dimensionless SDC which is a function of $m_c$, $m_b$ and the NRQCD factorization scale $\mu_{\Lambda}$.
Different from the normalization in Eq.~(\ref{eqn:dc full}), we have adopted the conventional nonrelativistic normalization for $V$ state in the long-distance matrix element (LDME) in Eq.~(\ref{eqn:fac}).
It is worth noting that the $\mu_{\Lambda}$ dependence in the SDC can be thoroughly canceled by that in the LDME.

The SDC can be determined by matching the renormalized vertex function in QCD and NRQCD, i.e.
\begin{align} \label{Master:formula}
&&Z_V\sqrt{Z_{2,b} Z_{2,c} } \,  \Gamma_{\rm QCD} =
\mathcal{C}(m_c, m_b, \mu_\Lambda) \, {\widetilde Z}_V^{-1}(\mu_\Lambda) \nonumber \\
&&\times\sqrt{\widetilde{Z}_{2,b} \widetilde{Z}_{2,c} }
\widetilde{\Gamma}_{\rm NRQCD} + {\mathcal O}(v^2),
\end{align}
where $\Gamma_{\rm QCD}$ ($\widetilde{\Gamma}_{\rm NRQCD}$) denotes the vertex function in QCD (NRQCD),
$Z_{2, Q}$ ($\widetilde{Z}_{2, Q}$) is the heavy quark on-shell field-strength renormalization constant in QCD (NRQCD),
and $Z_V$ ($\widetilde{Z}_V$) is the renormalization constant of the vector current in full QCD (NRQCD).
Since the NRQCD corrections only involve scaleless integrals, we can set $\widetilde{Z}_{2,Q}=1$ in dimensional regularization.
The value of $Z_V$ is given by $Z_V=1$.
It has been known for a long time that the $\mathcal{O}(\alpha_s)$ correction to $\widetilde{Z}_V$ vanishes,
so one can express $\widetilde{Z}_V$ up to $\mathcal{O}(\alpha_s^3)$ as
\begin{equation}
\widetilde{Z}_V = 1 + \left( \frac{\alpha_{s}^{(n_l)}(\mu_\Lambda)}{\pi}\right)^2 \delta\tilde{Z}^{(2)}+\left( \frac{\alpha_{s}^{(n_l)}(\mu_\Lambda)}{\pi}\right)^3 \delta\tilde{Z}^{(3)} + \mathcal{O}(\alpha_s^4),
\end{equation}
where we use $\alpha_{s}^{(n)}$ to denote the strong coupling constant for the effective field theory with $n$ active quarks.
One main purpose of this work is to determine the renormalization constants therein.

Another main goal of this work is to compute the SDC $\mathcal{C}(m_c,m_b,\mu_{\Lambda})$ up to $\mathcal{O}(\alpha_s^3)$ using the matching equation (\ref{Master:formula}).
In our calculation, we set the quantum number of the $c\bar{b}$ pair to be $^3S_1$ and employ the strategy of region~\cite{Beneke:1997zp}.
Specifically, before performing the loop integration in $\Gamma_{\rm QCD}$,
we neglect the relative momentum between the $c$ and $\bar{b}$ quarks,
which amounts to directly extracting the contribution from the hard region in the context of the strategy of region.
Starting at $\mathcal{O}(\alpha_s^2)$, the matching coefficient exhibits IR divergences,
which can be compensated by the UV divergences of the effective theory, making the SDC finite.
In Eq. (\ref{Master:formula}), the renormalization constant $\widetilde{Z}_V$ takes over this part.

In our computation, we include the contributions from the loops of charm quark and bottom quark,
which however are decoupled in the NRQCD effective field theory.
To keep consistency, we first compute the SDC by including contributions from the charm- and bottom-quark loops,
then obtain the SDC with $n_l$ active quarks by use of the
decoupling relation for $\alpha_s$~\cite{Grozin:2007fh,Chetyrkin:1997un}:
\begin{equation}
\begin{aligned}\label{decoupling-formula}
&\frac{\alpha_{s}^{(n_f+1)}(\mu)}{\pi}=\frac{\alpha_{s}^{(n_f)}(\mu)}{\pi} \\
&+\bigg(\frac{\alpha_{s}^{(n_f)}(\mu)}{\pi}\bigg)^{2}T_F\bigg[\frac{1}{3}\ln\frac{\mu^2}{m_Q^2}+(\frac{1}{6}\ln^2\frac{\mu^2}{m_Q^2}
+\frac{1}{36}\pi^2)\epsilon+\mathcal{O}(\epsilon^2)\bigg] \\
&+\bigg(\frac{\alpha_{s}^{(n_f)}(\mu)}{\pi}\bigg)^{3}T_F\bigg[(\frac{1}{4}\ln\frac{\mu^2}{m_Q^2}+\frac{15}{16})C_F
+(\frac{5}{12}\ln\frac{\mu^2}{m_Q^2}-\frac{2}{9})C_A \\
&+\frac{1}{9}T_F\ln^2\frac{\mu}{m_Q}+\mathcal{O}(\epsilon)\bigg]+\mathcal{O}(\alpha_s^4),
\end{aligned}
\end{equation}
where $m_Q$ denotes the mass of the decoupled heavy quark,
$T_F=1/2$, $C_F=(N_c^2-1)/(2N_c)$, $C_A=N_c$, and $N_c=3$ is the number of colors.

It is convenient to expand the dimensionless SDC $\mathcal{C}$ in powers of $\alpha_s$:
\begin{equation}
\begin{aligned}
\label{eqn:sdc:decomposition}
&\mathcal{C}(m_c,m_b,\mu_{\Lambda})=1+\frac{\alpha_s^{\left(n_l\right)}\left(\mu_R\right)}{\pi} \mathcal{C}^{(1)} \\
&~+\bigg(\frac{\alpha_s^{\left(n_l\right)}\left(\mu_R\right)}{\pi}\bigg)^2
\bigg(\mathcal{C}^{(1)}\frac{\beta_0}{4}\text{ln}\frac{\mu_{R}^2}{m_c m_b}+\gamma^{(2)}\ln \frac{\mu_{\Lambda}^2}{m_c m_b}+\mathcal{C}^{(2)}\bigg) \\
&~+\bigg(\frac{\alpha_s^{\left(n_l\right)}\left(\mu_R\right)}{\pi}\bigg)^3\bigg[\big(\frac{\mathcal{C}^{(1)}}{16}\beta_1+\frac{\mathcal{C}^{(2)}}{2}\beta_0\big)\text{ln}\frac{\mu_{R}^2}{m_c m_b} \\
&~+\frac{\mathcal{C}^{(1)}}{16}\beta^2_0 \ln^2
\frac{\mu_{R}^2}{m_c m_b}+\frac{1}{4}\big(2\frac{d\gamma^{(3)}}{d \text{ln}\mu_{\Lambda}^2}
-\beta_0\gamma^{(2)}\big)\ln^2\frac{\mu_{\Lambda}^2}{m_c m_b} \\
&~+\big(\mathcal{C}^{(1)}\gamma^{(2)}+\gamma^{(3)}\big)\ln\frac{\mu_{\Lambda}^2}{m_c m_b}+\frac{\beta_{0}}{2}\gamma^{(2)}\ln\frac{\mu_{\Lambda}^2}{m_c m_b}\,\text{ln}\frac{\mu_{R}^2}{m_c m_b} \\
&~+\mathcal{C}^{(3)}\bigg]+\mathcal{O}\left(\alpha_s^4\right),
\end{aligned}
\end{equation}
where $\mu_R$ is the QCD renormalization scale,
$\beta_0=(11/3)C_A-(4/3) T_F n_l$ and $\beta_1=(34/4)C_A^2-(20/3) C_A T_F n_l-4 C_F T_F n_l$ are the one-loop and two-loop $\beta$ functions.
The explicit form of the $\ln \mu_R$ and $\ln \mu_\Lambda$ terms are deduced from the renormalization group invariance and factorization scale invariance, respectively.
The anomalous dimensions $\gamma^{(2)}$ and $\gamma^{(3)}$ at $\mathcal{O}(\alpha_s^2)$ and $\mathcal{O}(\alpha_s^3)$ are associated with the NRQCD operator $\chi^{\dagger}_b{\bm \sigma}\psi_c$,
and are related to the renormalization constant $\widetilde{Z}_V$ by
\begin{equation}\label{eqn:gamma}
\begin{aligned}
&\gamma\left(r, {\mu^2_\Lambda\over m_c m_b } \right) \equiv
{d \ln \widetilde{Z}_V \over d \ln \mu_\Lambda^2 }
=\left(\frac{\alpha_s^{(n_l)}
\left(\mu_\Lambda\right)}{\pi}\right)^2 \gamma^{(2)}(r) \\
&~~~~~~+\left(\frac{\alpha_s^{\left(n_l\right)}\left(\mu_\Lambda\right)}{\pi}\right)^3\gamma^{(3)}
\left(r, {\mu^2_\Lambda\over m_c m_b}\right)+\mathcal{O}(\alpha^4_s).
\end{aligned}
\end{equation}
Since the anomalous dimension $\gamma$ is finite, we can deduce that
\begin{subequations}\label{eqn:gamma}
\begin{eqnarray}
\gamma^{(2)}&=&-2 {\widetilde Z}_V^{(2,-1)}, \\
\gamma^{(3)}&=&-3 {\widetilde Z}_V^{(3,-1)},
\end{eqnarray}
\end{subequations}
where ${\widetilde Z}_V^{(a,b)}$ denotes the coefficient of order $\alpha_s^a\epsilon^b$ in ${\widetilde Z}_V$.

$\mathcal{C}^{(1)}$, $\mathcal{C}^{(2)}$ and $\mathcal{C}^{(3)}$, as functions of $r$, correspond to the nontrivial parts of the SDC
at $\mathcal{O}(\alpha_s)$, $\mathcal{O}(\alpha^2_s)$, and $\mathcal{O}(\alpha^3_s)$, respectively.
Since the SDC is symmetric under $m_b\leftrightarrow m_c$, $\cal{C}$ as well as $\mathcal{C}^{(1)}$, $\mathcal{C}^{(2)}$ and $\mathcal{C}^{(3)}$ must be invariant
under $r\leftrightarrow 1/r$.

\section{Technicalities~\label{sec:tech}}

\begin{figure}[t]
\center{
\includegraphics*[scale=0.5]{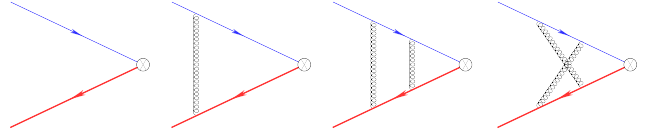}\qquad
\caption {\label{fig:bare} Representative Feynman diagrams for $c\bar{b}\to W$ through two-loop order.
The cross implies the insertion of the vector current.
}}
\end{figure}

\begin{figure}[t]
\center{
\includegraphics*[scale=0.5]{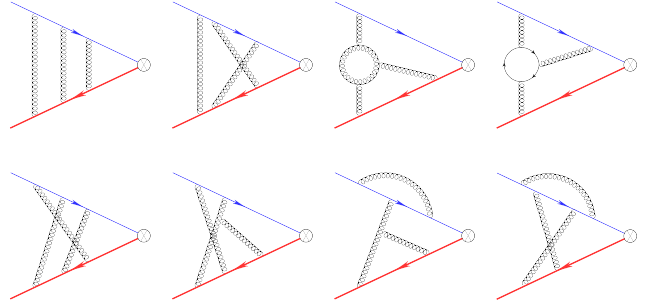}\qquad
\caption {\label{fig:bare3} Some typical Feynman diagrams for $c\bar{b}\to W$ at three-loop order. The cross implies the insertion of the vector current.
}}
\end{figure}

We use the package \texttt{FeynArts}~\cite{Hahn:2000kx} to generate the Feynman diagrams and amplitudes up to $\mathcal{O}(\alpha_s^3)$.
There are $1$, $1$, $14$, and $268$ Feynman diagrams for this process at tree, one-loop, two-loop, and three-loop orders, respectively.
Some representative Feynman diagrams are shown in Fig.\ref{fig:bare} and Fig.\ref{fig:bare3}.
We employ the covariant spin-triplet and color-singlet projector to construct the quarkonium decay amplitude.
The packages {\tt FeynCalc}\cite{Mertig:1990an} and {\tt FormLink}~\cite{Feng:2012tk} are used for the Dirac algebra calculations.

In the calculation beyond leading order in $\alpha_s$,
we employ the packages \texttt{Apart}~\cite{Feng:2012iq} to do partial fraction and \texttt{FIRE}~\cite{Smirnov:2014hma} to carry out the integration-by-parts (IBP) reduction.
We end up with $2$, $15$, $412$ master integrals (MIs) at one-loop, two-loop, and three-loop orders respectively.
The two one-loop MIs are simply two massive tadpoles, which can be easily evaluated.
All the analytical expressions of the appearing two-loop MIs can be directly evaluated or obtained by swapping the charm and bottom quarks in the results given in Ref.~\cite{Chen:2015csa},
which studied the the two-loop corrections to the decay width of $B_c\to l\nu$.
It is quite challenging to analytically evaluate the three-loop MIs, so we use high-precision numerical computation instead.
The recently released package AMFlow~\cite{Liu:2022chg},
which is based on the numerical differential equation algorithm known as the "Auxiliary Mass Flow" method~\cite{Liu:2017jxz,Liu:2020kpc,Liu:2021wks,Liu:2022mfb},
proves to be effective for tackling multi-loop MIs~\cite{Sang:2022kub,Zhang:2022nuf,Feng:2022vvk,Feng:2022ruy,Sang:2022erv,Chen:2022vzo,Chen:2022mre}.

In our computation, we work in Feynman gauge, and adopt dimensional regularization with the spacetime dimensions $D=4-2\epsilon$ to regularize both UV and IR divergences.
To eliminate the UV divergences, we implement the on-shell renormalization for the quark mass and field strength~\cite{Fael:2020bgs},
and the $\overline{\rm MS}$ renormalization for the QCD coupling constant.
In addition to the UV divergences, there exists an uncanceled single IR pole at two-loop order, and some uncanceled double and single IR poles at three-loop order.
As previously mentioned, these IR poles will be eliminated by the divergence in the renormalization constant $\widetilde{Z}_V$,
thus $\cal C$ in \eqref{Master:formula} is IR finite.

\section{SDC at one and two loops~\label{sec:two-loop}}
The one-loop QCD correction to $\cal C$, denoted by $\mathcal{C}^{(1)}(r)$ in \eqref{eqn:sdc:decomposition},
has a particularly simple structure~\cite{Hwang:1999fc}:
\begin{align}
& \mathcal{C}^{(1)}(r)=C_F \left[\frac{3(r-1)}{4(r+1)}\,\ln r-2\right].
\label{eqn:sdc trivial}
\end{align}

Next, we turn to computing the two-loop QCD correction to $\mathcal{C}$, i.e. $\mathcal{C}^{(2)}(r)$.
After using the IBP reduction to simplify the loop integrals, we obtain 15 two-loop MIs.
Among these, 3 MIs are simply products of two one-loop tadpole integrals,
10 MIs have been explicitly presented in Ref.\cite{Chen:2015csa}.
The remaining 2 MIs can also be obtained from Ref.\cite{Chen:2015csa} by making use of the symmetry between the charm quark and bottom quark.
With all the necessary ingredients in hand, we can now compute $\mathcal{C}^{(2)}(r)$ analytically.
At $\mathcal{O}(\alpha_s^2)$, the renormalized QCD perturbative calculation still contains an uncanceled single IR pole.
This is a common feature of NRQCD factorization that has been observed many times in high-order perturbative calculations involving quarkonium.
This IR pole should be canceled by the divergence in $\delta\tilde{Z}^{(2)}$, thus $\mathcal{C}^{(2)}(r)$ becomes IR finite.

Through computation, we obtain
\begin{equation}\label{eq:z-2}
\delta \tilde{Z}^{(2)} =  \pi ^2 C_F \frac{1}{\epsilon} \bigg[\frac{1}{8} C_A+\frac{C_F \left(3 r^2+2 r+3\right)}{24(r+1)^2}\bigg],
\end{equation}
and the corresponding anomalous dimension $\gamma^{(2)}$ using Eq.~(\ref{eqn:gamma}).
The result in Eq.~(\ref{eq:z-2}) is new,
and can be reduced to the corresponding results for the vector quarkonium $J/\psi$ or $\Upsilon$ decay in the limit of $r\to 1$.

The nontrivial $\mathcal{C}^{(2)}(r)$ is also obtained.
It is useful to organize $\mathcal{C}^{(2)}(r)$ according to its flavor and color structures as
\begin{equation}\label{eq:c2}
\mathcal{C}^{(2)}(r) = C_F^2 s_F+C_F C_A s_A+C_F T_F n_b s_b +C_F T_F n_c s_c +C_F T_F n_l s_l,
\end{equation} 	
where $n_c=1$ and $n_b=1$ denote the numbers of charm quark and bottom quark respectively,
and the coefficient $s_A$ is given by
\begin{equation}\label{eq:sA}
\begin{aligned}
&s_A=\frac{\pi ^2 \left(8 r^4-11 r^3+171 r^2+391 r+153\right)}{288 (r+1)^2}-2\zeta (3)-\frac{151}{72} \\
&-\frac{2}{3} \pi ^2 \ln 2-\frac{\pi ^2 \left(12 r^2-8 r-84\right)+183 (1-r)}{288 (r+1)}\ln r \\
&+\frac{\pi ^2 \left(3 r^2-14 r+3\right)}{48 r} \ln (r+1)+\frac{r \left(4 r^3+5 r^2+6 r+9\right)}{24 (r+1)^2} \ln^2 r \\
&+\frac{4 r^5+r^4+5 r^3-5 r^2-r-4}{24 r^2 (r+1)} H_{-,0}(r)-\frac{7 r^3-9 r^2+9 r-7}{48 r \left(r+1\right)} H_{+,0}(r) \\
&+\frac{3r^2-8r+3}{24r}\left( H_{0,-,0}(r)-H_{-,0,0}(r)\right)+\frac{1}{4} \left( H_{0,+,0}(r)-H_{+,0,0}(r)\right) ,
\end{aligned}
\end{equation}
$s_F$ given by
\begin{equation}\label{eq:sF}
\begin{aligned}
&s_F=-\frac{\pi ^2 \left(8 r^4-9 r^3+349 r^2+597 r+319\right)}{144(r+1)^2}-\frac{(r-1)^2 \zeta (3)}{2 (r+1)^2} \\
&+\frac{23}{8}+\frac{4}{3}\pi ^2 \ln 2+\frac{8 \pi ^2 \left(3 r^3+13 r^2+5 r+3\right)-525 \left(r^2-1\right)}{288 (r+1)^2} \ln r \\
&-\frac{32 r^4+36 r^3-3 r^2+30 r-27}{96 (r+1)^2} \ln^2 r \\
&-\frac{8 r^5+r^4+5 r^3-5 r^2-r-8}{24 r^2 (r+1)} H_{-,0}(r) \\
&-\frac{\pi ^2 \left(3 r^4+10 r^3+6 r^2+10 r+3\right)}{24 r (r+1)^2} \ln (r+1) \\
&+\frac{3 r^3-7 r^2+7 r-3}{12 r (r+1)} H_{+,0}(r) \\
&-\frac{3 r^4+7 r^3+4 r^2+7 r+3}{12 r (r+1)^2} \left( H_{0,-,0}(r)-H_{-,0,0}(r)\right) \\
&+\frac{3 r^2+2 r+3}{12 (r+1)^2} \left( H_{0,+,0}(r)-H_{+,0,0}(r)\right) ,
\end{aligned}
\end{equation}
$s_b$ given by
\begin{equation}\label{eq:sb}
\begin{aligned}
&s_b=r^2+\frac{\pi ^2 \left(36 r^5-71 r^4-54 r^3-118 r+15\right)}{288(r+1)}-\frac{13 r}{24}+\frac{143}{72} \\
&+\frac{15 r^3+4 r^2+2 r+21}{24 (r+1)} \ln r+\frac{9 r^5+7 r^4-3 r+3}{12 (r+1)} \left(\ln^2 r+H_{-,0}(r)\right) \\
&-\frac{33 r^4+18 r^3+22 r-9}{48 (r+1)} H_{+,0}(r) ,
\end{aligned}
\end{equation}
$s_c$ given by
\begin{equation}\label{eq:sc}
\begin{aligned}
&s_c=\frac{1}{r^2}-\frac{\pi ^2 (3 r+5)}{36 (r+1)}-\frac{13}{24r}+\frac{143}{72}-\frac{21 r^3+2 r^2+4 r+15}{24 r^2 (r+1)} \ln r \\
&-\frac{3 r^5-3 r^4+7 r+9}{12 r^4 (r+1)} H_{-,0}(r)-\frac{9 r^4-22 r^3-18 r-33}{48 r^3 (r+1)} H_{+,0}(r),
\end{aligned}
\end{equation}
and $s_l$ given by
\begin{equation}\label{eq:sL}
s_L = \frac{11}{18}-\frac{5 (r-1)}{24 (r+1)}\ln r.
\end{equation}
Here $H$ denotes the harmonic polylogarithms (HPLs).
For the reference of readers, we list all the appearing HPLs in terms of logarithms and polylogarithms:
\begin{equation}
\begin{aligned}
&H_{\pm,0}(r)  = \text{Li}_2\left(\frac{1}{r}\right)\pm\text{Li}_2(-r)+\frac{1}{2}\ln^2 r-\ln (r-1) \ln r \\
&~~~~\pm \ln (r+1) \ln r-\frac{\pi ^2}{3} ,\\
&H_{0,\pm,0}(r)=-2 \text{Li}_3\left(\frac{1}{r}\right)\pm 2\text{Li}_3(-r)-\text{Li}_2\left(\frac{1}{r}\right) \ln r \\
&~~~~\mp \text{Li}_2(-r) \ln r -\frac{1}{6} \ln^3 r-\frac{1}{3} \pi ^2\ln r , \\
&H_{\pm,0,0}(r)  = \text{Li}_3\left(\frac{1}{r}\right)\mp\text{Li}_3(-r)+\text{Li}_2\left(\frac{1}{r}\right) \ln r\pm \text{Li}_2(-r) \ln r \\
&~~~~+\frac{1}{3}\ln^3 r-\frac{1}{2} \ln (r-1) \ln^2 r\pm \frac{1}{2} \ln (r+1) \ln^2 r.
\end{aligned}
\end{equation}

It is worth noting that the expressions of Eqs.~(\ref{eq:sA}-\ref{eq:sL}) are valid only for $r>1$.
We can obtain the analytical expressions of $\mathcal{C}^{(2)}$ for $r<1$ by applying the symmetry under $r\leftrightarrow 1/r$,
i.e. $\mathcal{C}^{(2)}(r)=\mathcal{C}^{(2)}(1/r)$.
As a check, we verify that $\mathcal{C}^{(2)}(r)$ in the limit of $r\to 1$ agrees with the corresponding expression for the decay constant of $J/\psi$~\cite{Czarnecki:1997vz,Beneke:1997jm}.

\section{SDC at three loops~\label{sec:three-loop}}

After IBP reduction for the three-loop QCD perturbative amplitude, we end up with 412 MIs.
It is much more challenging to deduce the analytical expressions for the encountered three-loop MIs.
In this work, we are satisfied with obtaining high-precision numerical results for a series of values of $r$. Similar to the case at two loops,
after renormalization, the QCD amplitude still contains uncanceled double and single IR divergences,
which should be eliminated by the divergence in $\delta\tilde{Z}^{(3)}$.

Since the renormalization factor $\delta \widetilde{Z}^{(3)}$ is a function of $r$ rather than a constant,
reconstructing its analytical expression is somewhat challenging.
Our strategy is to first compute the QCD amplitude with very high numerical accuracy at several different values of $r$,
then use the \texttt{PSLQ} algorithm~\cite{ferguson1999analysis} to convert the numerical results to the corresponding rational numbers,
and finally apply Thiele's interpolation formula~\cite{abramowitz1964handbook} to reconstruct $\delta\widetilde{Z}^{(3)}(r)$.
After some efforts, we obtain the final results
\begin{equation}\label{eqn:z3}
\begin{aligned}
&\delta \widetilde{Z}^{(3)} =  \pi^2 C_F \bigg\lbrace \frac{1}{\epsilon^2} \left( \frac{3 r^2-r+3}{36 (r+1)^2} C_F^2+\frac{r}{216 (r+1)^2} C_F C_A-\frac{1}{16} C_A^2 \right) \\
&+\frac{1}{\epsilon} \bigg[ \bigg(  \frac{19 r^2+5 r+19}{36 (r+1)^2}+\frac{1}{6} \ln (r+1)-\frac{2}{3}\ln 2 \\
&+\frac{\left(r^3-4 r^2-2 r-3\right)}{12 (r+1)^3}\ln r+\frac{\left(3 r^2-r+3\right) }{12 (r+1)^2}\ln \frac{\mu_\Lambda ^2}{m_c m_b}  \bigg) C_F^2 \\
&+\bigg( \frac{1}{4} \ln (r+1) -\frac{(r+11) }{48 (r+1)}\ln r  +\frac{39 r^2+148 r+39}{162 (r+1)^2} \\
&+\frac{\left(11 r^2+8 r+11\right) }{48 (r+1)^2}\ln \frac{\mu_\Lambda ^2}{m_c m_b} \bigg) C_F C_A \\
&+\bigg(  \frac{1}{12} \ln (r+1)-\frac{1}{24}\ln r  +\frac{1}{6}\ln 2+\frac{2}{27}+\frac{1}{24} \ln \frac{\mu_\Lambda ^2}{m_c m_b}\bigg)C_A^2 \bigg] \\
&+T_F n_b \frac{1}{15  (r+1)^2}\frac{1}{\epsilon}C_F+T_F n_c \frac{ r^2}{15 (r+1)^2}\frac{1}{\epsilon} C_F \\
&+T_F n_l \bigg[ \bigg( \frac{3 r^2+2 r+3}{108 (r+1)^2} \frac{1}{\epsilon^2}-\frac{21 r^2+58 r+21}{324 (r+1)^2} \frac{1}{\epsilon}\bigg)C_F \\
&+\left( \frac{1}{36} \frac{1}{\epsilon^2}-\frac{37}{432}\frac{1}{\epsilon} \right)C_A \bigg] \bigg\rbrace .
\end{aligned}
\end{equation}
The expression for $\delta\widetilde{Z}^{\left(3\right)}$ is complicated and is known for the first time.
A new feature is that $\delta\widetilde{Z}^{\left(3\right)}$ also explicitly depends on the factorization scale $\mu_\Lambda$.
One can easily verify that $\delta\widetilde{Z}^{(3)}$ is indeed symmetric under the exchanges $r\leftrightarrow 1/r$ and $m_c\leftrightarrow m_b$.
Furthermore, in the $r\to 1$ limit, $\delta\widetilde{Z}^{(3)}$ exactly reproduces the expression of $\widetilde{Z}_v$ in Ref.~\cite{Egner:2022jot},
the renormalization constant associated with the vector NRQCD current with equal quark masses.

By using the relation \eqref{eqn:gamma},
we can extract the anomalous dimension $\gamma^{(3)}$ for the NRQCD operator $\chi^{\dagger}_b{\bm \sigma}\psi_c$ from Eq.~(\ref{eqn:z3}).
Compared to the two-loop anomalous dimension $\gamma^{(2)}$,
the three-loop one has a more complicated expression and depends on $\mu_\Lambda$.

Now, the only missing piece in the three-loop SDC in \eqref{eqn:sdc:decomposition} is $\mathcal{C}^{(3)}(r)$,
which is independent of $\mu_R$ and $\mu_\Lambda$, and only depends on $r$.
Following the convention of Refs.~\cite{Marquard:2014pea,Beneke:2014qea,Egner:2022jot,Feng:2022vvk},
we find it convenient to decompose $\mathcal{C}^{(3)}(r)$ in terms of different color and flavor structures:
\begin{equation}
\begin{aligned}
&\mathcal{C}^{(3)}(r)=C_F\Big\{ C^2_F\mathcal{C}_{FFF}+C_FC_A\mathcal{C}_{FFA}+C_A^2\mathcal{C}_{FAA} \\
&+T_F n_l\bigg[C_F\mathcal{C}_{FFL} +C_A\mathcal{C}_{FAL}+T_F\big(n_c\mathcal{C}_{FCL}+n_b\mathcal{C}_{FBL}+n_l\mathcal{C}_{FLL}\big)\bigg] \\
&+T_F^2n_bn_c\mathcal{C}_{FBC}+T_Fn_c\big[C_F\mathcal{C}_{FFC}+C_A\mathcal{C}_{FAC}+T_Fn_c\mathcal{C}_{FCC}\big] \\
&+T_Fn_b\big[C_F\mathcal{C}_{FFB}+ C_A\mathcal{C}_{FAB}+T_Fn_b\mathcal{C}_{FBB}\big] \Big\}.
\label{eqn:C_col_str}
\end{aligned}
\end{equation}

To make numerical computations, we choose the values of charm and bottom masses to be $m_c=2.04$ GeV, and $m_{b}=4.98$ GeV,
which correspond to the three-loop heavy quark pole masses,
converted from the known $\overline{\rm MS}$ masses $\overline{m}_c(\overline{m}_c)=1.28$ GeV and $\overline{m}_b(\overline{m}_b)= 4.18$ GeV~\cite{Workman:2022ynf} using the three-loop formula~\cite{Herren:2017osy}.
With this choice, $r=4.98/2.04\approx2.44$, and we present the numerical values for the coefficients of the various color/flavor structures in (\ref{eqn:C_col_str}):
\begin{subequations}\label{c3rphys}
\begin{eqnarray}
 \mathcal{C}_{FAA} &=& -100.25110229918804820, \\ 
\mathcal{C}_{FFA} &=& -195.31280533643102048, \\ 
\mathcal{C}_{FAB} &=& -0.20449501444189092375, \\ 
 \mathcal{C}_{FAC} &=& 0.27747159652902911973, \\ 
 \mathcal{C}_{FAL} &=& 39.950899279146777293, \\ 
 \mathcal{C}_{FFF} &=& 27.987972624700185294, \\ 
 \mathcal{C}_{FFB}&=& -0.23586009689621660478, \\ 
 \mathcal{C}_{FFC} &=& -1.5659590509720943966, \\ 
 \mathcal{C}_{FFL} &=& 49.308208345142224459, \\ 
 \mathcal{C}_{FBB} &=& 0.020486039573427998510, \\ 
 \mathcal{C}_{FBC}&=& 0.094676866596467921217, \\ 
 \mathcal{C}_{FBL} &=& -0.084564878004863959208, \\ 
 \mathcal{C}_{FCC} &=& 0.12700615294260801711, \\ 
 \mathcal{C}_{FCL} &=& -0.63407618072083251486, \\ 
 \mathcal{C}_{FLL} &=& -2.2296659667275518411.  
\end{eqnarray}
\end{subequations}

From the values in Eq.~(\ref{c3rphys}), we see that the terms proportional to $C_F C_A^2$ and $C_F^2 C_A$ are the most and second-most important,
while the contributions from the charm and bottom quark loops are quite small.

It is instructive to assess the importance of the radiative corrections.
By using the aforementioned heavy quark masses, setting the renormalization scale $\mu_R=\sqrt{m_b m_c}$,
and choosing the factorization scale $\mu_\Lambda=1$ GeV, which corresponds to the typical energy scale $m_Q v_Q$, we have
\begin{align}\label{C:pert:expansion-1}
&\mathcal{C}=1-2.29\left(\frac{\alpha_s^{\left(n_l\right)}}{\pi}\right)-35.36\left(\frac{\alpha_s^{\left(n_l\right)}}{\pi}\right)^2 \nonumber \\
&~~~~~~-1686.80\left(\frac{\alpha_s^{\left(n_l\right)}}{\pi}\right)^3+\mathcal{O}(\alpha_s^4),
\end{align}
for $n_l=3, n_c=n_b=1$,
\begin{align}\label{C:pert:expansion-2}
&\mathcal{C}=1-2.29\left(\frac{\alpha_s^{\left(n_l\right)}}{\pi}\right)-35.44\left(\frac{\alpha_s^{\left(n_l\right)}}{\pi}\right)^2 \nonumber \\
&~~~~~~-1686.27\left(\frac{\alpha_s^{\left(n_l\right)}}{\pi}\right)^3+\mathcal{O}(\alpha_s^4),
\end{align}
for $n_l=3, n_c=1, n_b=0$,
\begin{align}\label{C:pert:expansion-3}
&\mathcal{C}=1-2.29\left(\frac{\alpha_s^{\left(n_l\right)}}{\pi}\right)-35.77\left(\frac{\alpha_s^{\left(n_l\right)}}{\pi}\right)^2 \nonumber \\
&~~~~~~-1686.63\left(\frac{\alpha_s^{\left(n_l\right)}}{\pi}\right)^3+\mathcal{O}(\alpha_s^4),
\end{align}
for $n_l=3, n_c=n_b=0$, and
\begin{align}\label{C:pert:expansion-3}
&\mathcal{C}=1-2.29\left(\frac{\alpha_s^{\left(n_l\right)}}{\pi}\right)-36.83\left(\frac{\alpha_s^{\left(n_l\right)}}{\pi}\right)^2 \nonumber \\
&~~~~~~-2026.07\left(\frac{\alpha_s^{\left(n_l\right)}}{\pi}\right)^3+\mathcal{O}(\alpha_s^4),
\end{align}
for $n_l=n_c=n_b=0$.
It is clear that the contributions from the heavy quark loop are negligible,
and the contribution from the light quark loop is small at two loops but becomes significant at three loops.

To check our results, we take $r=1.0001$ and obtain
\begin{equation}\label{C:c3:asy}
\begin{aligned}
&\mathcal{C}(r=1.0001)=1-\left(\frac{\alpha_s^{\left(n_l\right)}}{\pi}\right)2.66667 \\
&~~~~+\left(\frac{\alpha_s^{\left(n_l\right)}}{\pi}\right)^2\bigg(-44.55101+0.40741 n_l\bigg) \\
&~~~~+\left(\frac{\alpha_s^{\left(n_l\right)}}{\pi}\right)^3\bigg(-2090.33289+120.66107n_l-0.82278n_l^2\bigg) \\
&~~~~+\mathcal{O}(\alpha_s^4),
\end{aligned}
\end{equation}
where we have set $n_c=1$, $n_b=0$, and $\mu_R=\mu_\Lambda=\sqrt{m_c m_b}$.
We find our result in Eq.~(\ref{C:c3:asy}) is consistent with the value of $c_v$ in Eq.~(17) in Ref.~\cite{Egner:2022jot},
which corresponds to the SDC of the vector NRQCD current in the equal quark mass case.

\begin{figure}[t]
\center{
\includegraphics*[scale=0.55]{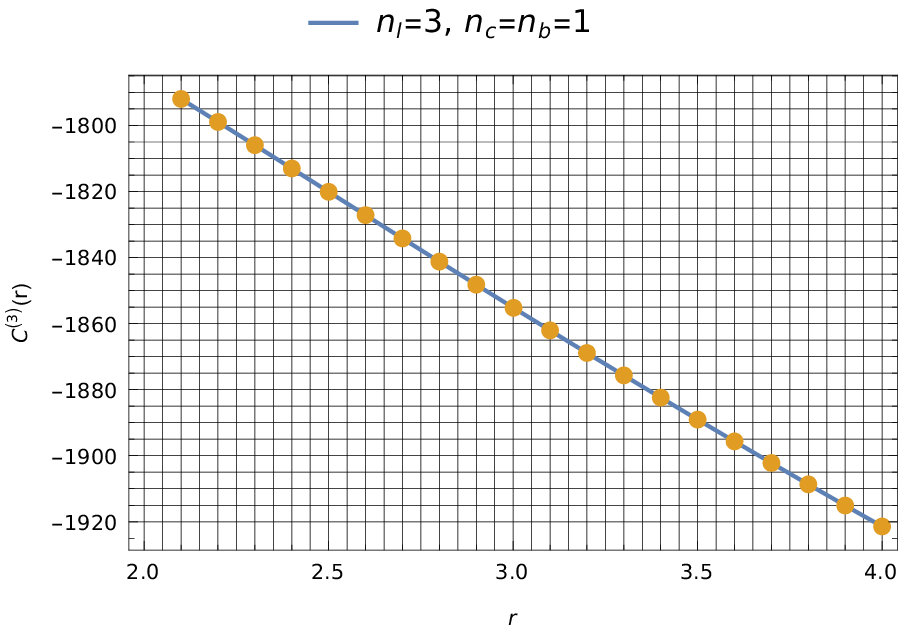}\qquad
\includegraphics*[scale=0.55]{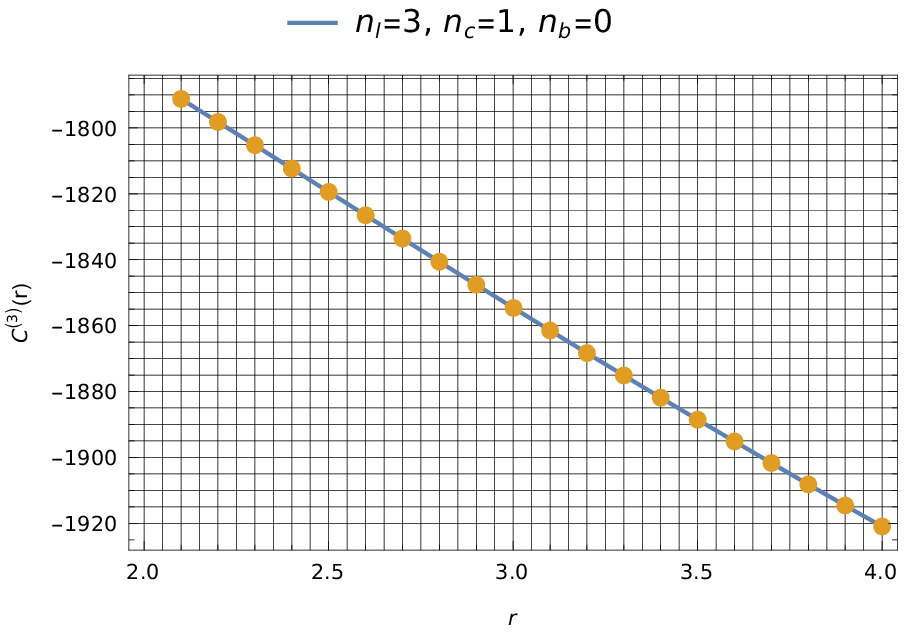}\qquad
\includegraphics*[scale=0.55]{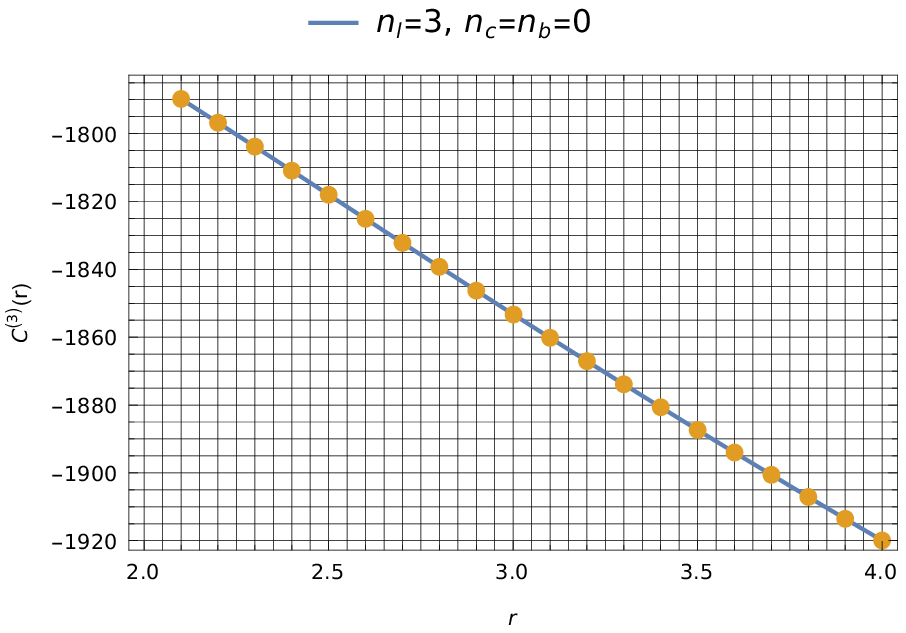}\qquad
\includegraphics*[scale=0.55]{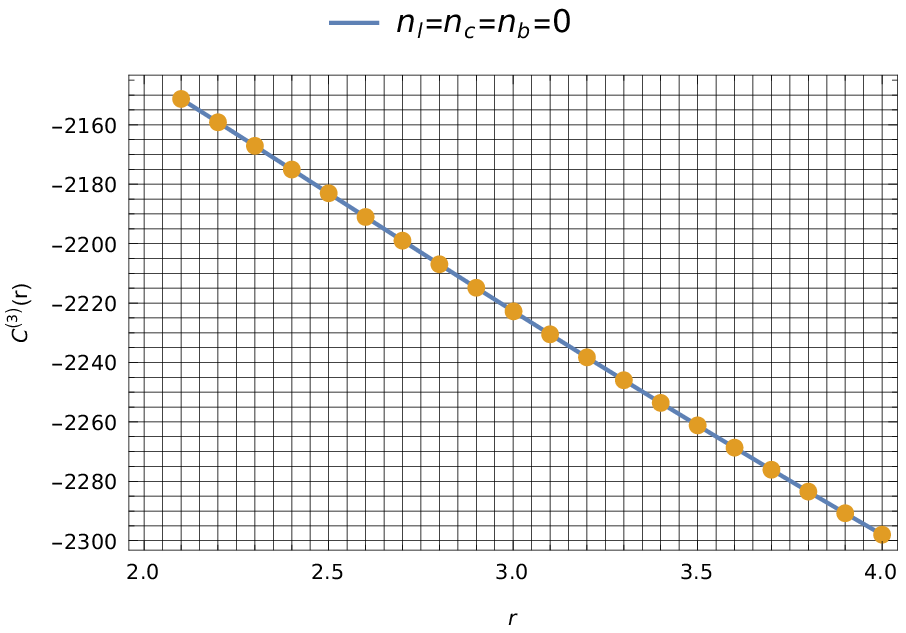}\qquad
\caption {\label{fig:c-r} The dimensionless SDC $\mathcal{C}^{(3)}(r)$ as a function of $r$ at various choice of $n_l$, $n_c$ and $n_b$.
}}
\end{figure}

It is worth exploring the dependence of $\mathcal{C}^{(3)}(r)$ on $r$ further.
Although it is challenging to derive the analytical expression of $\mathcal{C}^{(3)}(r)$ at this stage,
we can numerically compute this function for a range of values of $r$.
After some work, we obtain numerical results for $r$ ranging from $2.1$ to $4.0$,
with an interval of $0.1$. The explicit results are presented in Tabs.~\ref{tab:c3-1} and \ref{tab:c3-2} in the appendix.
Additionally, we plot $\mathcal{C}^{(3)}(r)$ as a function of the mass ratio $r$ for different choices of $n_l$, $n_c$, and $n_b$ in Fig.\ref{fig:c-r}.
From these figures, we are surprised to find that the shape of the function $\mathcal{C}^{(3)}(r)$ is approximately linear with respect to $r$ in the range $(2.1,4.0)$,
i.e., $\mathcal{C}^{(3)}(r)\approx -1649.17-68.42 r$ for $n_l=3$ and $n_c=n_b=1$,
and $\mathcal{C}^{(3)}(r)\approx -1989.35-77.53 r$ for $n_l=n_c=n_b=0$.
The root-mean-square errors for these two fits are 0.66 and 0.69, respectively.
The reason for this behavior is unclear and warrants further investigation.

\section{phenomenology~\label{sec:phy}}
Employing the values of the decay constant of $B_c^*$,
we can easily evaluate its leptonic decay width using the following equation,
\begin{equation}\label{eq:decay-rate}
\Gamma[B_c^*\to \ell \nu_\ell]=\frac{1}{12\pi}|V_{bc}|^2 G_F^2 m_{V}^3 (1-\frac{m_\ell^2}{m_V^2})^2(1+\frac{m_\ell^2}{2m_V^2})|f_V|^2,
\end{equation}
where $\ell$ can be any of the charged lepton, $e, \nu, \tau $, $m_\ell$ is the mass of $\ell$,
$V_{bc}$ and $G_F$ denote the Cabibbo-Kobayashi-Maskawa matrix element and the Fermi coupling constant of the weak interaction, respectively.

In the numerical analysis, we need to fix the values of the various input parameters.
The values of the following parameters are taken from the latest particle data group~\cite{ParticleDataGroup:2022pth}:
\begin{equation}\label{eq:para}
\begin{aligned}
&m_e = 0.000510999\,  {\rm GeV}, \\
&m_\mu = 0.10566\,  {\rm GeV}, \\
&m_\tau = 1.777\, {\rm GeV}, \\
&V_{bc}=0.0408, \\
&G_F=1.16638\times 10^{-5}\, {\rm GeV}^{-2}.
\end{aligned}
\end{equation}
In addition, we take the heavy quark pole masses $m_c=2.04$ GeV and $m_{b}=4.98$ GeV. In this choice,  $r=4.98/2.04\approx2.44$. 
Since $B_c^*$ has not been observed in experiment,
we take $m_V=6.3$ GeV from Ref.~\cite{Eichten:1994gt} computed by the Buchm\"uller-Tye (BT) potential model.
We use Eq.(\ref{eqn:fac}) to calculate $f_V$,
which is the product of the SDC given in Eq.(\ref{eqn:sdc:decomposition})
and the NRQCD LDME $\langle 0 |\chi^{\dagger}_b\bm{\sigma}\cdot \bm{\epsilon}\psi_c(\mu_{\Lambda})|V \rangle$.
The LDME is actually dependent on the factorization scale, and its evolution is given by the following equation,
\begin{equation}\label{eq:evolution}
\begin{aligned}
&\frac{d\langle 0 |\chi^{\dagger}_b\bm{\sigma}\cdot \bm{\epsilon}\psi_c(\mu_{\Lambda})|V \rangle}{d\ln\mu^2_\Lambda}=-\left(\frac{\alpha_s^{(n_l)}
\left(\mu_\Lambda\right)}{\pi}\right)^2 \gamma^{(2)} \\
&~~~~~~-\left(\frac{\alpha_s^{\left(n_l\right)}\left(\mu_\Lambda\right)}{\pi}\right)^3\gamma^{(3)}+\mathcal{O}(\alpha^4_s).
\end{aligned}
\end{equation}
Currently, there is no well-established treatment for this LDME at $\mathcal{O}(\alpha_s^3)$ in theory.
For our phenomenological prediction, we will approximate its value at $\mu_\Lambda=1\, \rm GeV$ by solving the Schr\"odinger equation,
in which the potential is given by the BT potential model~\cite{Eichten:1995ch}, i.e.
\begin{equation}\label{eq:evolution}
|\langle 0 |\chi^{\dagger}_b\bm{\sigma}\cdot \bm{\epsilon}\psi_c|V \rangle|^2\approx
\frac{N_c}{2\pi}|R^{c\bar{b}}_{1S}(0)|^2=\frac{3}{2\pi}\times 1.642\, {\rm GeV}^3.
\end{equation}

Using the ingredients described above, we present our predictions for the decay widths at various levels of accuracy in Tab.~\ref{tab:1}.
From the table, we can see that all the perturbative corrections to the decay constant are negative,
and the $\mathcal{O}(\alpha_s)$, $\mathcal{O}(\alpha_s^2)$, and $\mathcal{O}(\alpha_s^3)$ corrections are small, considerable, and very significant, respectively.
This indicates that the perturbative convergence is very poor. When we include all the corrections,
we find that the theoretical value for $f_V$ is negative at $\mu_R=\sqrt{m_c m_b}$, which poses a disturbing puzzle to theory.
In fact, one can obtain a positive prediction for the decay constant by taking a larger value for $\mu_R$,
but the result is sensitive to the value of $\mu_R$ due to the large perturbative corrections.
This difficulty deserves further study.
Finally, we want to emphasize that it is straightforward to predict the decay constant
and the decay width using different values of $m_c$ and $m_b$ with the formulas provided in this work.

\begin{table*}[!htbp]\small
\caption{Theoretical prediction on the decay widths at various levels of accuracy.
The labels LO, NLO, $\rm N^2LO$ and $\rm N^3LO$ indicate the results accurate up to leading order,
$\mathcal{O}(\alpha_s)$, $\mathcal{O}(\alpha_s^2)$, and $\mathcal{O}(\alpha_s^3)$ respectively.
We take $m_c=2.04$ GeV and $m_b=4.98$ GeV, $m_V=6.3$  GeV, $\mu_R=\sqrt{m_c m_b}$, and $\mu_\Lambda=1$ GeV.}
	\label{tab:1}
	\setlength{\tabcolsep}{0.4mm}
	\centering
{
	\begin{tabular}{|c|c|c|c|c|}
	\hline
	& LO & NLO & $\rm N^2LO$ & $\rm N^3LO$ \\
	\hline
$f_V$ (GeV)	& \qquad 0.499\qquad\qquad  & \qquad 0.408 \qquad\qquad & \qquad 0.298  \qquad\qquad  &\qquad -0.119 \qquad\qquad \\
  \hline
$\Gamma(B_c^*\to e \nu_e)\times 10^{4}$ (eV)  & \qquad 3.738 \qquad\qquad  & \qquad 2.505 \qquad\qquad & \qquad 1.334 \qquad\qquad  & \qquad 0.212 \qquad\qquad \\
	\hline
$\Gamma(B_c^*\to \mu \nu_\mu)\times 10^{4}$ (eV)  &\qquad 3.737 \qquad \qquad &  \qquad 2.504 \qquad \qquad &  \qquad 1.333 \qquad \qquad & \qquad 0.212 \qquad\qquad \\
	\hline
$\Gamma(B_c^*\to \tau \nu_\tau)\times 10^{4}$ (eV)  &\qquad 3.293 \qquad\qquad & \qquad 2.207 \qquad\qquad  & \qquad 1.175  \qquad\qquad & \qquad 0.186 \qquad\qquad \\
	\hline
	\end{tabular}
}
\end{table*}

\section{Summary~\label{sec:sum}}
In summary, we carry out two-loop and three-loop calculations of the decay constant of the $B_c^*$ particle within the framework of the NRQCD formalism.
This is the first time that the analytical expressions for the renormalization constant and the corresponding anomalous dimension associated with the NRQCD vector current composed of $c\bar{b}$ have been obtained at $\mathcal{O}(\alpha_s^2)$ and $\mathcal{O}(\alpha_s^3)$.
These expressions are functions of $m_c$ and $m_b$, as well as the factorization scale $\mu_\Lambda$.
We also compute the SDC up to $\mathcal{O}(\alpha_s^3)$ in perturbative expansion.
Specifically, we derive the analytical expressions for the $\mathcal{O}(\alpha_s^2)$ corrections
in terms of logarithmic and polylogarithmic functions of the ratio $r=m_c/m_b$,
and we have provided numerical results for the $\mathcal{O}(\alpha_s^3)$ corrections at a range of values of $r$.
Interestingly, we find that the nontrivial $\mathcal{O}(\alpha_s^3)$ SDC can be accurately approximated by a simple linear function of $r$.
The phenomenological predictions is also explored.
We find that the $\mathcal{O}(\alpha_s^2)$ and $\mathcal{O}(\alpha_s^3)$ corrections to the decay constant and decay width are considerable and very significant,
which indicates a very poor convergence of the perturbative expansion.

\vspace*{0.25cm}
{\bf Acknowledgements.}
We thank L.-B. Chen for providing us the analytic expressions of MIs ${\rm I}8$, ${\rm I}9$, ${\rm I}10$ and ${\rm I}11$ in Ref.~\cite{Chen:2015csa}.
The work of W.-L. S. and M.-Z. Z is supported by the National Natural Science Foundation of China
under Grants No. 11975187. The work of H.-F. Zhang is supported by
the National Natural Science Foundation of China under Grants No. 11965006.

\vspace*{0.25cm}
{\bf Note added.} - When preparing our manuscript,
we noticed an independent preprint~\cite{Tao:2022qxa} about the same topic, where the authors
numerically compute the $\mathcal{O}(\alpha_s^2)$ corrections to the decay constants of $B_c$ and $B_c^*$.  By taking the same parameters, we find our $\mathcal{O}(\alpha_s^2)$ result for the SDCs of $B_c^*$ is consistent with that in Ref.~\cite{Tao:2022qxa}. In addition, we 
distinguish the contributions from the charm-quark loop and bottom-quark loop, while it is not in Ref.~\cite{Tao:2022qxa}.

\section*{Appdendix}
\appendix
\section{Numerical results of the SDC at various values of $r$~\label{sec:app}}
In this appendix, we present the numerical expressions for the nontrivial parts of the SDC
$\mathcal{C}^{(1)}$, $\mathcal{C}^{(2)}$, and $\mathcal{C}^{(3)}$ with $r$ ranging from $2.1$ to $4.0$ with a fixed step $0.1$ in Tabs.~\ref{tab:c3-1}~\footnote{Since coefficients of some MIs are divergent at $r=3.0$, we approximate the
result at $r=3.0$ by that at $r=2.9998$.}
and \ref{tab:c3-2}.
For the reference of readers, we keep the explicit $n_l$, $n_c$ and $n_b$ dependence.

\begin{table*}[!htbp]\small
	\caption{The results of $\mathcal{C}^{(1)}$, $\mathcal{C}^{(2)}$, and $\mathcal{C}^{(3)}$ at various values of $r$.}
	\label{tab:c3-1}
	\setlength{\tabcolsep}{0.4mm}
	\centering
{
	\begin{tabular}{|c|c|c|c|}
		\hline
		{$r$}
		& $\mathcal{C}^{(1)}$
		& $\mathcal{C}^{(2)}$
		& $\mathcal{C}^{(3)}$  \\
		\hline
		\multirow{2}*{$2.1$}
		& \multirow{2}*{$-2.40$}
		& $-47.37+0.37 n_l$
		&        $-2151.29+122.81 n_l - 0.93 n_c - 0.68 n_b-0.77 n_l^2$                   \\
		&
		&$+ 0.29 n_c+0.09 n_b$
		&$+0.04 n_c^2+0.01 n_b^2-0.19 n_l n_c-0.04 n_l n_b+0.03 n_c n_b$ \\
		\hline
		\multirow{2}*{$2.2$}
		& \multirow{2}*{$-2.37$}
		& $-47.70+0.37 n_l$
		&        $-2159.15+123.08 n_l - 0.91 n_c - 0.66 n_b-0.76 n_l^2$                   \\
		&
		&$+ 0.30 n_c+0.09 n_b$
		&$+0.04 n_c^2+0.01 n_b^2-0.19 n_l n_c-0.03 n_l n_b+0.03 n_c n_b$ \\
		\hline
		\multirow{2}*{$2.3$}
		& \multirow{2}*{$-2.34$}
		& $-48.02+0.36 n_l$
		&        $-2167.08+123.35 n_l - 0.88 n_c - 0.64 n_b-0.75 n_l^2$                   \\
		&
		&$+ 0.31 n_c+0.08 n_b$
		&$+0.04 n_c^2+0.01 n_b^2-0.20 n_l n_c-0.03 n_l n_b+0.03 n_c n_b$ \\
		\hline
		\multirow{2}*{$2.4$}
		& \multirow{2}*{$-2.31$}
		& $-48.35+0.36 n_l$
		&        $-2175.05+123.62 n_l - 0.85 n_c - 0.63 n_b-0.75 n_l^2$                   \\
		&
		&$+ 0.32 n_c+0.08 n_b$
		&$+0.04 n_c^2+0.01 n_b^2-0.21 n_l n_c-0.03 n_l n_b+0.03 n_c n_b$ \\
		\hline
		\multirow{2}*{$2.5$}
		& \multirow{2}*{$-2.27$}
		& $-48.67+0.35 n_l$
		&        $-2183.04+123.89 n_l - 0.82 n_c - 0.61 n_b-0.74 n_l^2$                   \\
		&
		&$+ 0.33 n_c+0.08 n_b$
		&$+0.04 n_c^2+0.01 n_b^2-0.22 n_l n_c-0.03 n_l n_b+0.03 n_c n_b$ \\
		\hline
		\multirow{2}*{$2.6$}
		& \multirow{2}*{$-2.24$}
		& $-48.99 +0.35 n_l$
		&        $-2191.03 + 124.17 n_l - 0.78 n_c - 0.60 n_b-0.73 n_l^2$                   \\
		&
		&$+ 0.34 n_c+0.07 n_b$
		&$+0.05 n_c^2+0.01 n_b^2-0.22 n_l n_c-0.03 n_l n_b+0.03 n_c n_b$ \\
		\hline
		\multirow{2}*{$2.7$}
		& \multirow{2}*{$-2.21$}
		& $-49.30+0.34 n_l$
		&        $-2199.00+124.44 n_l - 0.75 n_c - 0.58 n_b-0.73 n_l^2$                   \\
		&
		&$+ 0.35 n_c+0.07 n_b$
		&$+0.05 n_c^2+0.01 n_b^2-0.23 n_l n_c-0.02 n_l n_b+0.03 n_c n_b$ \\
		\hline
		\multirow{2}*{$2.8$}
		& \multirow{2}*{$-2.18$}
		& $-49.61+0.34 n_l$
		&        $-2206.94+124.72 n_l - 0.71 n_c - 0.57 n_b-0.72 n_l^2$                   \\
		&
		&$+ 0.36 n_c+0.07 n_b$
		&$+0.05 n_c^2+0.01 n_b^2-0.24 n_l n_c-0.02 n_l n_b+0.03 n_c n_b$ \\
		\hline
		\multirow{2}*{$2.9$}
		& \multirow{2}*{$-2.15$}
		& $-49.92+0.34 n_l$
		&        $-2214.85 + 125.00 n_l - 0.68 n_c - 0.56 n_b-0.71 n_l^2$                   \\
		&
		&$+ 0.37 n_c+0.07 n_b$
		&$+0.05 n_c^2+0.01 n_b^2-0.24 n_l n_c-0.02 n_l n_b+0.03 n_c n_b$ \\
		\hline
		\multirow{2}*{$3.0$}
		& \multirow{2}*{$-2.12$}
		& $-50.22 + 0.33 n_l$
		&        $-2222.7 + 125.27 n_l - 0.64 n_c - 0.55 n_b-0.71 n_l^2$                   \\
		&
		&$+ 0.38 n_c+0.07 n_b$
		&$+0.05 n_c^2+0.01 n_b^2-0.25 n_l n_c-0.02 n_l n_b+0.03 n_c n_b$ \\
		\hline
	\end{tabular}
}
\end{table*}

\begin{table*}[!htbp]\small
	\caption{The results of $\mathcal{C}^{(1)}$, $\mathcal{C}^{(2)}$, and $\mathcal{C}^{(3)}$ at various values of $r$.}
	\label{tab:c3-2}
	\setlength{\tabcolsep}{0.4mm}
	\centering
{
	\begin{tabular}{|c|c|c|c|}
		\hline
		{$r$}
		& $\mathcal{C}^{(1)}$
		& $\mathcal{C}^{(2)}$
		& $\mathcal{C}^{(3)}$  \\
		\hline
		\multirow{2}*{$3.1$}
		& \multirow{2}*{$-2.09$}
		& $-50.52+0.33 n_l$
		&        $-2230.52+ 125.55 n_l - 0.60 n_c - 0.53 n_b-0.70 n_l^2$                   \\
		&
		&$+ 0.39 n_c+0.06 n_b$
		&$+0.05 n_c^2+0.01 n_b^2-0.25 n_l n_c-0.02 n_l n_b+0.03 n_c n_b$ \\
		\hline
		\multirow{2}*{$3.2$}
		& \multirow{2}*{$-2.06$}
		& $-50.82 +0.32 n_l$
		&        $-2238.28 + 125.82 n_l - 0.55 n_c - 0.52 n_b-0.69 n_l^2$                   \\
		&
		&$+ 0.39 n_c+0.06 n_b$
		&$+0.06 n_c^2+0.01 n_b^2-0.26 n_l n_c-0.02 n_l n_b+0.03 n_c n_b$ \\
		\hline
		\multirow{2}*{$3.3$}
		& \multirow{2}*{$-2.03$}
		& $-51.11+0.32 n_l$
		&        $-2245.98 + 126.09 n_l - 0.51 n_c - 0.51 n_b-0.69 n_l^2$                   \\
		&
		&$+ 0.40 n_c+0.06 n_b$
		&$+0.06 n_c^2+0.01 n_b^2-0.27 n_l n_c-0.02 n_l n_b+0.03 n_c n_b$ \\
		\hline
		\multirow{2}*{$3.4$}
		& \multirow{2}*{$-2.00$}
		& $-51.39 + 0.31 n_l$
		&        $-2253.61 + 126.36 n_l - 0.47 n_c - 0.50 n_b-0.68 n_l^2$                   \\
		&
		&$+ 0.41 n_c+0.06 n_b$
		&$+0.06 n_c^2+0.01 n_b^2-0.27 n_l n_c-0.02 n_l n_b+0.03 n_c n_b$ \\
		\hline
		\multirow{2}*{$3.5$}
		& \multirow{2}*{$-1.97$}
		& $-51.67+0.31 n_l$
		&        $-2261.17 + 126.63 n_l - 0.43 n_c - 0.49 n_b-0.68 n_l^2$                   \\
		&
		&$+ 0.42 n_c+0.06 n_b$
		&$+0.06 n_c^2+0.01 n_b^2-0.28 n_l n_c-0.02 n_l n_b+0.03 n_c n_b$ \\
		\hline
		\multirow{2}*{$3.6$}
		& \multirow{2}*{$-1.94$}
		& $-51.95 +0.31 n_l$
		&        $-2268.67 + 126.90 n_l - 0.38 n_c - 0.49 n_b-0.67 n_l^2$                   \\
		&
		&$+ 0.42 n_c+0.06 n_b$
		&$+0.06 n_c^2+0.01 n_b^2-0.28 n_l n_c-0.01 n_l n_b+0.03 n_c n_b$ \\
		\hline
		\multirow{2}*{$3.7$}
		& \multirow{2}*{$-1.92$}
		& $-52.22 + 0.30 n_l$
		&        $-2276.10 + 127.17 n_l - 0.34 n_c - 0.48 n_b-0.66 n_l^2$                   \\
		&
		&$+ 0.43 n_c+0.06 n_b$
		&$+0.07 n_c^2+0.01 n_b^2-0.29 n_l n_c-0.01 n_l n_b+0.03 n_c n_b$ \\
		\hline
		\multirow{2}*{$3.8$}
		& \multirow{2}*{$-1.89$}
		& $-52.49 + 0.30 n_l$
		&        $-2283.46 + 127.43 n_l - 0.29 n_c - 0.47 n_b-0.66 n_l^2$                   \\
		&
		&$+ 0.44 n_c+0.05 n_b$
		&$+0.07 n_c^2 + 0.01 n_b^2-0.30 n_l n_c-0.01 n_l n_b+0.03 n_c n_b$ \\
		\hline
		\multirow{2}*{$3.9$}
		& \multirow{2}*{$-1.86$}
		& $-52.75 + 0.30 n_l$
		&        $-2290.75 + 127.70 n_l - 0.24 n_c - 0.46 n_b-0.65 n_l^2$                   \\
		&
		&$+ 0.45 n_c+ 0.05 n_b$
		&$+0.07 n_c^2+0.01 n_b^2-0.30 n_l n_c-0.01 n_l n_b+0.03 n_c n_b$ \\
		\hline
\multirow{2}*{$4.0$}
		& \multirow{2}*{$-1.83$}
		& $-53.01 + 0.29 n_l$
		&        $-2297.97 + 127.96 n_l - 0.20 n_c - 0.46 n_b-0.65 n_l^2$                   \\
		&
		&$+ 0.45 n_c+0.05 n_b$
		&$+0.07 n_c^2+0.01 n_b^2-0.31 n_l n_c-0.01 n_l n_b+0.03 n_c n_b$ \\
		\hline
	\end{tabular}
}
\end{table*}


\end{document}